\documentclass[useAMS,usenatbib,usegraphicx]{mn2e}

\usepackage{graphicx}
\usepackage[latin1]{inputenc}
\usepackage{color}
\usepackage{times}
\usepackage{natbib}
\usepackage{setspace}
\newif\ifAMStwofonts
\AMStwofontstrue
\definecolor{red}{rgb}{1,0.,0.}

\newcommand{\msun}{{\rm M}_\odot}
\newcommand{\msunyr}{{\rm M}_\odot\ {\rm yr}^{-1}}
\newcommand{\gaea}{\sc{gaea}}
\def\lesssim{\lower.5ex\hbox{$\; \buildrel < \over \sim \;$}}
\def\gtrsim{\lower.5ex\hbox{$\; \buildrel > \over \sim \;$}}
\title[CR-IGIMF] {On the shape and evolution of a cosmic ray regulated
  galaxy-wide stellar initial mass function.} \author[Fontanot et al.]{
  \parbox[t]{\textwidth}{Fabio Fontanot$^1$\thanks{E-mail:
      fabio.fontanot@inaf.it}, Francesco La Barbera$^2$, Gabriella De Lucia$^1$, Anna Pasquali$^3$, \\
    Alexandre Vazdekis$^{4,5}$}
    \vspace*{8pt}\\
    $^1$ INAF - Astronomical Observatory of Trieste, via G.B. Tiepolo 11, I-34143 Trieste, Italy \\
    $^2$ INAF - Astronomical Observatory  of Capodimonte, sal. Moiariello, 16, I-80131, Napoli, Italy \\
    $^3$ Astronomisches Rechen-Institut, Zentrum f\"ur Astronomie, Universit\"at Heidelberg, M\"onchhofstr. 12-14, D-69120 Heidelberg, Germany \\
    $^4$ Instituto de Astrof\'isica de Canarias, E-38200 La Laguna, Tenerife, Spain \\
    $^5$ Departamento de Astrof\'isica, Universidad de La Laguna, E-38205 La Laguna, Tenerife, Spain\\
}

\begin{document}
\date{Accepted ... Received ...}

\maketitle

\begin{abstract} 
In this paper, we present a new derivation of the shape and evolution
of the integrated galaxy-wide initial mass function (IGIMF),
incorporating explicitly the effects of cosmic rays (CRs) as
regulators of the chemical and thermal state of the gas in the dense
cores of molecular clouds. We predict the shape of the IGIMF as a
function of star formation rate (SFR) and CR density, and show that it
can be significantly different with respect to local estimates. In
particular, we focus on the physical conditions corresponding to IGIMF
shapes that are {\it simultaneously} shallower at high-mass end {\it
  and} steeper at the low-mass end than a Kroupa IMF. These solutions
can explain both the levels of $\alpha$-enrichment and the excess of
low-mass stars as a function of stellar mass, observed for local
spheroidal galaxies. As a preliminary test of our scenario, we use
idealized star formation histories to estimate the mean IMF shape for
galaxies of different $z=0$ stellar mass. We show that the fraction of
low-mass stars as a function of galaxy stellar mass predicted by these
mean IMFs agrees with the values derived from high-resolution
spectroscopic surveys.
\end{abstract}


\section{Introduction}\label{sec:intro}                                       
In recent years, a wealth of observations have challenged the notion
of a universal stellar initial mass function (IMF) in early-type
galaxies (ETGs), pointing to possible variations as a function of
galaxy properties such as stellar mass ($M_\star$) or velocity
dispersion ($\sigma$).  Dynamical studies \citep[see e.g.][]{Treu10,
  Cappellari12, Dutton13} suggest a systematic excess of the
mass-to-light ratios derived using integral field stellar kinematics
with respect to the values estimated from photometry assuming a
Milky-Way (MW) -like (e.g. Kroupa or Chabrier) IMF. The excess is
found to increase with $\sigma$. It is important to stress that the
dynamical analysis is not able to disentangle \citep[see
  e.g.][]{Tortora16} between a 'top-heavy' and a 'bottom-heavy'
scenario, where the excess of stellar mass is due to an enhanced
fraction of either stellar remnants (from exploding Supernovae) or of
low-mass stars.

Spectroscopy can provide an alternative and more specific approach to
study IMF variations, by taking advantage of spectral features, such
as, e.g., the Na\,I doublet at
$\lambda\lambda8183,8195$\,\AA\ \citep[hereafter
  NaI8200]{FaberFrench80, Schiavon97b}, the TiO1 and
TiO2~\citep{Trager98} features ($\lambda\lambda6000,6300$\,\AA\ ), and
the Wing-Ford FeH band \citep{WingFord69, Schiavon97a} at
$9,900$\,\AA. These features depend on the IMF, because of their
sensitivity to stellar surface gravity and/or effective
temperature\footnote{Some features, like NaI8200, are specifically
  sensitive to surface gravity and thus to the IMF shape, while other
  indices, like TiO2, increase with decreasing stellar effective
  temperature. Therefore, their IMF sensitivity is the result of the
  increase of the number of low-mass (low effective temperature) stars
  relative to high-mass stars.}, and bear information on the ratio
between the total number of dwarf ($\rm m_\star \lesssim 0.5
M_{\odot}$) and more massive stars ($\rm m_\star \gtrsim 0.5
M_{\odot}$). In order to extract this information from high
signal-to-noise ratio spectra, a comparison with predictions from
stellar population synthesis models is required. Using this technique,
several studies have reported an excess of low- relative to high-mass
stars in the IMF of massive ETGs, i.e. a bottom-heavy IMF. The slope
at the low-mass end becomes increasingly steeper (in some cases
``super''-Salpeter, i.e.  with a dwarf-to-giant ratio exceeding that
of a Salpeter IMF) at increasing galaxy velocity dispersion or stellar
mass~\citep{Cenarro03, ConroyvanDokkum12, Ferreras13, LaBarbera13,
  Spiniello14a}. In addition, radial IMF gradients in local ETGs have
been measured~\citep{MartinNavarro15, LaBarbera17, vanDokkum17,
  Sarzi17}, suggesting that the dwarf-enhanced population is mostly
confined to the galaxy innermost regions (but see also
\citealt{Alton17}). It is worth noting that for old systems, such as
ETGs, the study of unresolved stellar populations through their
integrated light is sensitive only to long-lived stars, i.e. it does
not constrain the high-mass end slope of the IMF (i.e. above $\rm \sim
1 M_{\odot}$).  Moreover, IMF-sensitive features are mostly sensitive
to the dwarf-to-giant ratio in the IMF \citep[hereafter
  LB13]{LaBarbera13}, although the IMF shape might be constrained in
detail, by combining features at different
wavelengths~\citep{ConroyvanDokkum12, LaBarbera16, Lyubenova16}.
  
Possible IMF variations as a function of galaxy SFR activity have been
often proposed in the literature. Based on the analysis of optical
colours and $H_\alpha$ line-strengths for galaxies in the Galaxy And
Mass Assembly (GAMA) survey, \citet{Gunawardhana11} found evidence for
a flatter IMF {\it high-mass} slope in systems with higher star
formation rate. From a theoretical point of view, strongly
star-forming regions are expected to have the largest deviations with
respect to a universal, MW-like, IMF \citep[see
  e.g.][]{Klessen05}. \citet{WeidnerKroupa05} proposed a derivation of
the integrated galaxy-wide stellar IMF (IGIMF), based on a limited
number of physically and observationally motivated axioms. It is
possible to reformulate each of these as a function of SFR, and thus
predict the IMF shape as a function of this key physical property of
galaxies. The original IGIMF approach postulates the universality of
the IMF inside individual molecular clouds (MCs), with the variability
being driven by additional assumptions on the distribution of
molecular clouds in the galaxy and by physical considerations on the
mass of the most massive star that can form in a given MC. These
assumptions translate into an invariant shape for the low-mass end of
the IGIMF, which is at variance with results obtained from the spectra
of ETGs (but see Jerabkova et al. in preparation, for a recent update
of the IGIMF framework). A different approach has been taken by
\citet[][PP11 hereafter]{Papadopoulos11}, who consider the role of
cosmic rays (CRs), associated with SNe and stellar winds, in
regulating star formation in MCs. CRs are assumed to be very effective
in altering the thermal and chemical properties of the inner - UV
shielded - regions of MCs, eventually changing the relation between
gas density and temperature in MC cores. The characteristic Jeans mass
of young stars ($M^{\star}_{\rm J}$) forming in each MC is thus
affected. PP11 discuss numerical solutions to the thermal and chemical
equations describing the evolution of the interstellar medium in this
scenario (see also \citealt{Thi09}). These numerical solutions provide
an estimate for $M^{\star}_{\rm J}$ as a function of cluster core
density $\rho_{\rm cl}$ and CR ionization rate.

Although featuring a different evolution of the IMF shape (in the
IGIMF framework the evolution is mainly tied to the high-mass end,
whereas in the CR approach to the knee of the mass function), both
models predict similar variations in strongly star-forming objects,
i.e the IMF should become 'top-heavier', at the high mass end, than in
the local neighbourhood at increasing SFR or CR energy density. In
previous work, we tested the impact of the IGIMF \citep[F17
  hereafter]{Fontanot17a} and CR regulation \citep[F18
  hereafter]{Fontanot18} on the evolution of the physical and chemical
properties of galaxies, as predicted by the semi-analytic model
{\gaea} (GAlaxy Evolution and Assembly - \citealt{Hirschmann16}). We
have shown that in both frameworks we reproduce a mass and
mass-to-light ratio excess \citep{Cappellari12, Conroy13} with respect
to a MW-like IMF, which we interpret as driven by the mismatch between
intrinsic physical properties and those derived from synthetic
photometry under the assumption of a universal IMF. In addition, in
both scenarios we are able to reproduce, at the same time, the
observed increase for the [$\alpha$/Fe] ratio as a function of galaxy
stellar mass, and the mass-metallicity relation of ETGs.  Matching
both observables represents a known problem for hierarchical models
implementing a universal IMF~\citep{DeLucia17}. In particular,
reproducing the [$\alpha$/Fe]-$M_\star$ relation requires a balancing
between the relative number of Type-II and Type-Ia supernovae,
contributing to most of the $\alpha$-elements and Iron,
respectively. Assuming a 'top-heavy' IMF at increasing SFR helps
matching the observations, as it implies a larger fraction of Type-II
SNe in strong starbursts (associated with massive galaxies). However,
that is in contrast with the finding of a 'bottom-heavy' IMF in
massive galaxies. On the other hand, a 'bottom-heavy' IMF with a slope
steeper than Salpeter over the entire stellar mass range implies a low
fraction of Type-II supernovae and this is in contrast with the
observed metallicity and abundance ratios of massive galaxies. These
considerations are even more puzzling in light of recent studies
finding IMF radial trends in ETGs, with a steepening of the IMF
low-mass end slope in the innermost, more metal-rich, galaxy core
regions. In order to reconcile these apparently contradicting results,
\citet{Weidner13b} and \citet[see also
  \citealt{Vazdekis96}]{Ferreras15} have proposed a time-varying
scenario, where the low- and high-mass ends of the IMF vary
independently over different time scales, with the top-heavy regime
dominating the first (bursty) phase of star formation, followed by a
bottom-heavy regime. The implications of such a variable IMF for ETG
chemical evolution have been explored, e.g., in
\citet{DeMasi18}. However, a physical explanation for this
time-varying scenario, possibly related to the complex physics of gas
fragmentation in the dense core regions of massive galaxies at high
redshift, is still lacking.

In the present work, we combine the IGIMF and PP11 approaches to
derive a new formulation for the IGIMF that takes into account
explicitly the effect of CR heating on the IMF of individual MCs. We
present this new combined derivation, that we dubbed CR-IGIMF, in
Sec.~\ref{sec:compimf}, we show its basic properties in
Sec.~\ref{sec:props}, and we then discuss our results and their
implications in Section~\ref{sec:final}.

\section{Cosmic Rays regulated IGIMF}\label{sec:compimf}
In this paper, we present a derivation of the IGIMF ($\varphi_{\rm
  IGIMF}$) based on an approach similar to \citet[see also
  \citealt{Kroupa13} for a review]{WeidnerKroupa05}. We integrate the
IMF associated with individual clouds ($\varphi_\star(m)$), weighted
by the mass function of individual MCs ($\varphi_{\rm CL}(M_{\rm
  cl})$):

\begin{equation}
\varphi_{\rm IGIMF}(m) = \int^{M_{\rm cl}^{\rm max}}_{M_{\rm cl}^{\rm
    min}} \varphi_\star(m \le m_\star^{\rm max} (M_{\rm cl}))
\varphi_{\rm CL}(M_{\rm cl}) dM_{\rm cl}
\label{eq:igimf}
\end{equation}

The key quantities involved in the integration are the maximum value
of the mass of a star cluster ($M_{\rm cl}^{\rm max}$) and the largest
stellar mass ($m_{\rm max}$) forming in a given cluster. We set the
mass of the smallest star cluster to $M_{\rm cl}^{\rm min}=5 M_\odot$
following evidences from the Taurus-Auriga complex
\citep{KroupaBouvier03}. As in F17, $M_{\rm cl}^{\rm max}$ and $m_{\rm
  max}$ can be defined as a function of the instantaneous SFR using
the following axioms:

\begin{equation}
\log M_{\rm cl}^{\rm max} = 0.746 \log SFR+4.93;
\label{eq:mclmax}
\end{equation}

\begin{equation}
\begin{array}{ll}
\log m_\star^{\rm max} =& 2.56 \, \log M_{\rm cl} \times \\
    & [ \, 3.82^{9.17} + (\log M_{\rm cl})^{9.17} \, ]^{1/9.17} -0.38;\\
\label{eq:msmax}
\end{array}
\end{equation}

\begin{equation}
\varphi_{\rm CL}(M_{\rm cl}) \propto M_{\rm cl}^{-\beta};
\label{eq:clmf}
\end{equation}

\begin{equation}
\beta = \left\{
\begin{array}{ll}
2 & SFR < 1 M_\odot/yr \\
-1.06 \log SFR +2 & SFR \ge 1 M_\odot/yr \\
\end{array}
\right.
\label{eq:beta}
\end{equation}

\begin{equation}
\alpha_3 = \left\{
\begin{array}{ll}
2.35 & \rho_{\rm cl} < 9.5 \times 10^4 M_\odot/pc^3\\
1.86-0.43 \log(\frac{\rho_{\rm cl}}{10^4}) & \rho_{\rm cl} \ge 9.5 \times 10^4 M_\odot/pc^3 \\
\end{array}
\right.
\label{eq:alpha3}
\end{equation}

\begin{equation}
  \log \rho_{\rm cl} = 0.61 \log M_{\rm cl} +2.85.
\label{eq:rhocl}
\end{equation}

\noindent
Using stellar cluster data, \citet{Weidner04} derived
Eq.~\ref{eq:mclmax} to describe the dependence of $M_{\rm cl}^{\rm
  max}$ on the instantaneous SFR (see also \citealt{Kroupa13} for an
analytic derivation). As in F17, we impose\footnote{This value is
  consistent with the idea that present Globular Clusters might have
  had up to a factor 100 higher mass at birth
  \citep{Weidner04}. Whereas the exact value clearly depends on the
  IGIMF, we choose a rather conservative value. It is worth stressing
  that Eq.~\ref{eq:mclmax} predicts values larger than $M_{\rm
    cl}^{\rm max}$ only for SFR $\ge 10^{3.5} \msunyr$. In F17 and
  F18, we show that such strong SF events are not typical in
  theoretical models implementing a variable IMF.}  $M_{\rm cl}^{\rm
  max} \leq 2 \times 10^7 \msun$. Eq.~\ref{eq:msmax} represents a fit
to the numerical solution to the problem of finding the maximum
stellar mass forming in a cluster of mass $M_{\rm cl}$
\citep{PflammAltenburg07}, under the hypothesis that it contains
exactly one $m_{\rm max}$ star and using the canonical
IMF. Eq.~\ref{eq:clmf} describes the assumed functional shape for the
star cluster mass function. Allowed values for $\beta$
(Eq.~\ref{eq:beta}) are chosen based on local observations that
suggest $\beta=2$ \citep{LadaLada03}, and on $z\lesssim0.35$ data from
the GAMA survey \citep{Gunawardhana11} that require a flattening of
$\beta$ at high-SFRs. Eq.~\ref{eq:alpha3} takes into account possible
variations of the high-mass end slope $\alpha_3$ from the reference
choice $\alpha_3=2.35$ \citep[see][for a review]{Kroupa13}. As in F17,
in this work we assume $\alpha_3$ depends on $\rho_{\rm cl}$, as
proposed by \citet{Marks12}. This relation has the advantage of being
independent of other parameters (such as metallicity), that have been
shown to also correlate with $\alpha_3$ \citep[see
  e.g.][]{MartinNavarro15}. Finally, Eq.~\ref{eq:rhocl} describes the
relation between $\rho_{\rm cl}$ and $M_{\rm cl}$ as derived by
\citet{MarksKroupa12}.

Using the equations above, and assuming a given $\varphi_\star(m)$ for
individual MCs, it is then possible to construct the IGIMF
corresponding to an individual SF event. In the original
\citet{WeidnerKroupa05} framework, the IMF associated with individual
MCs has a canonical broken power-law shape with three slopes (as in
\citealt{Kroupa01}; see Eq.~1 in F17). In this paper, we use a
slightly different approach by considering a variable inner break at
$m_{\rm br}$:

\begin{equation}
\varphi_\star(m) =  \left\{
\begin{array}{ll}
(\frac{m}{m_{\rm low}})^{-\alpha_1} & m_{\rm low} \le m < m_{\rm br} \\
(\frac{m_{\rm br}}{m_{\rm low}})^{-\alpha_1} (\frac{m}{m_{\rm br}})^{-\alpha_2} & m_{\rm br} \le m < m_1 \\
(\frac{m_{\rm br}}{m_{\rm low}})^{-\alpha_1} (\frac{m_1}{m_{\rm br}})^{-\alpha_2} (\frac{m}{m_1})^{-\alpha_3} & m_1 \le m \le m_{\rm max} \\
\end{array}
\right.
\label{eq:kroimf}
\end{equation}

\noindent
where the parameters $m_{\rm low}=0.1$, $m_1=1.0$, $\alpha_1=1.3$ and
$\alpha_2=2.35$ are fixed. The position of the break at $m_{\rm br}$
is derived using the approach of PP11, assuming that it corresponds to
the characteristic Jeans mass ($M^{\star}_{\rm J}$) of young stars in
a MC with core density given by Eq.~\ref{eq:rhocl}, and affected by
the CR energy density $U_{\rm CR}$. Whereas in F18 we fixed the knee
to the value of $M^{\star}_{\rm J}$ for a typical density $\rho_{\rm
  cl}=10^5 cm^{-3}$, in the present paper we use a different approach,
considering a grid of values

\begin{equation}
m_{\rm br} = M^{\star}_{\rm J}(\rho_{\rm cl},U_{\rm CR}).
\end{equation}

For practical reasons, we use the numerical solutions to the chemical
and thermal equations for the CR regulated ISM presented in Fig.~4 of
PP11. These numerical experiments provide indeed interesting insights
on the key physical dependencies. In particular, $m_{\rm br}$
increases with $U_{\rm CR}$: this is mainly due to the higher CR
heating associated with the increased energy density. At fixed $U_{\rm
  CR}$, $m_{\rm br}$ is predicted to decrease with $\rho_{\rm
  cl}$. This behaviour is not trivial, as high density regions are
expected to be generally hotter, and the Jeans mass increases with
temperature and decreases with density. Hence, in the numerical
treatment of PP11, the dependence on temperature appears sub-dominant
with respect to that on density. The effect is important in our
analysis, as it allows changes in the low-mass end slope of the IGIMF
(as discussed below). In the following, we use the same notation as in
PP11, and consider $U_{\rm CR}$ values relative to the corresponding
MW one ($U_{\rm MW}$).

\section{Properties of the CR-IGIMF}\label{sec:props}
\begin{figure*}
  \centerline{ \includegraphics[width=18cm]{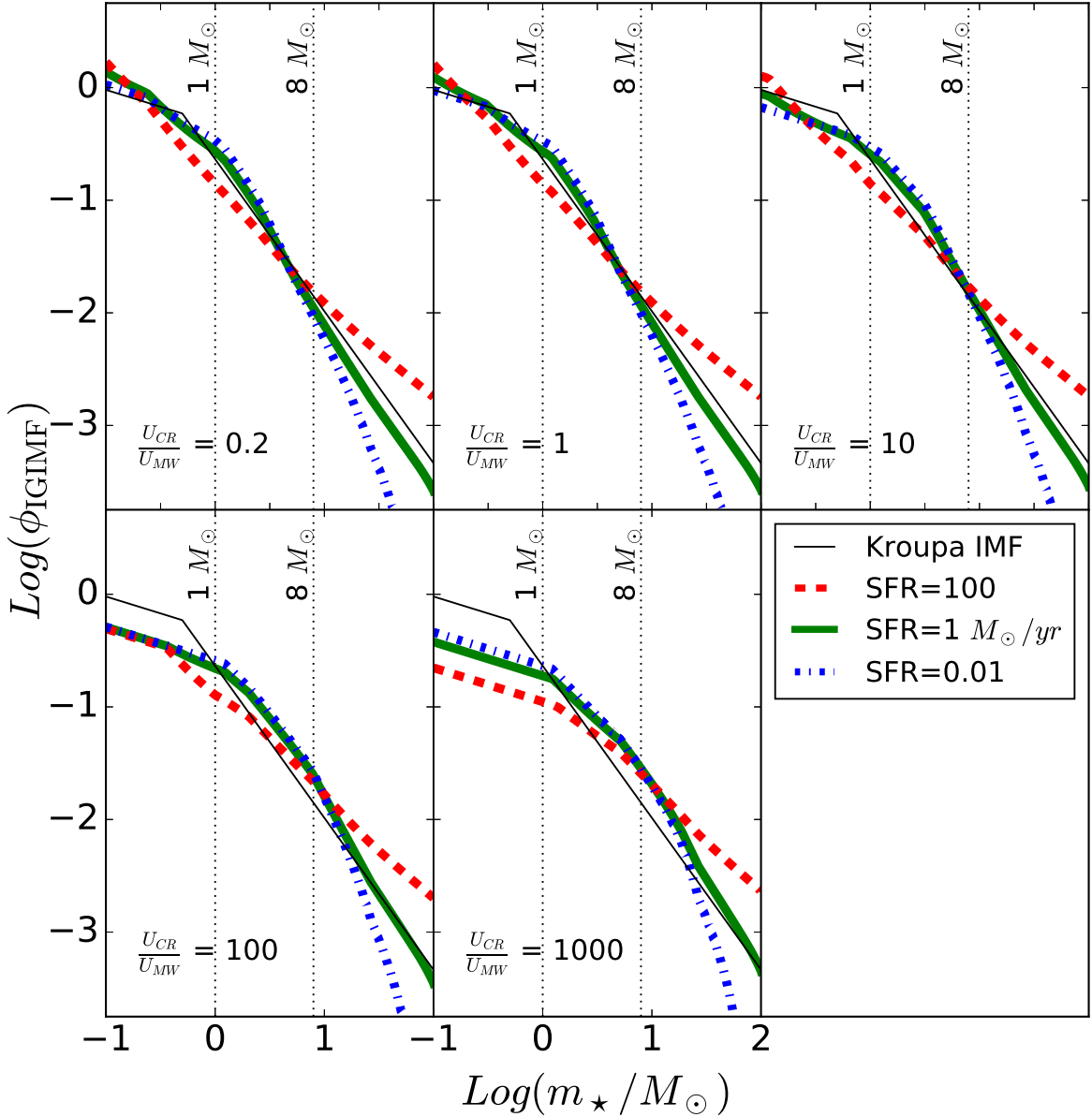} }
  \caption{Composite shapes for the Cosmic Ray regulated
    IGIMF. Different panels refer to different CR energy densities, as
    labelled in the lower--left corner of each panel. Red dashed,
    green solid and blue dotted lines correspond to SFR=100, 1, and
    0.01 $\msunyr$, respectively. In all panels, the thin solid line
    shows the canonical \citet{Kroupa01} IMF.}\label{fig:compimf}
\end{figure*}
\begin{figure}
  \centerline{ \includegraphics[width=9cm]{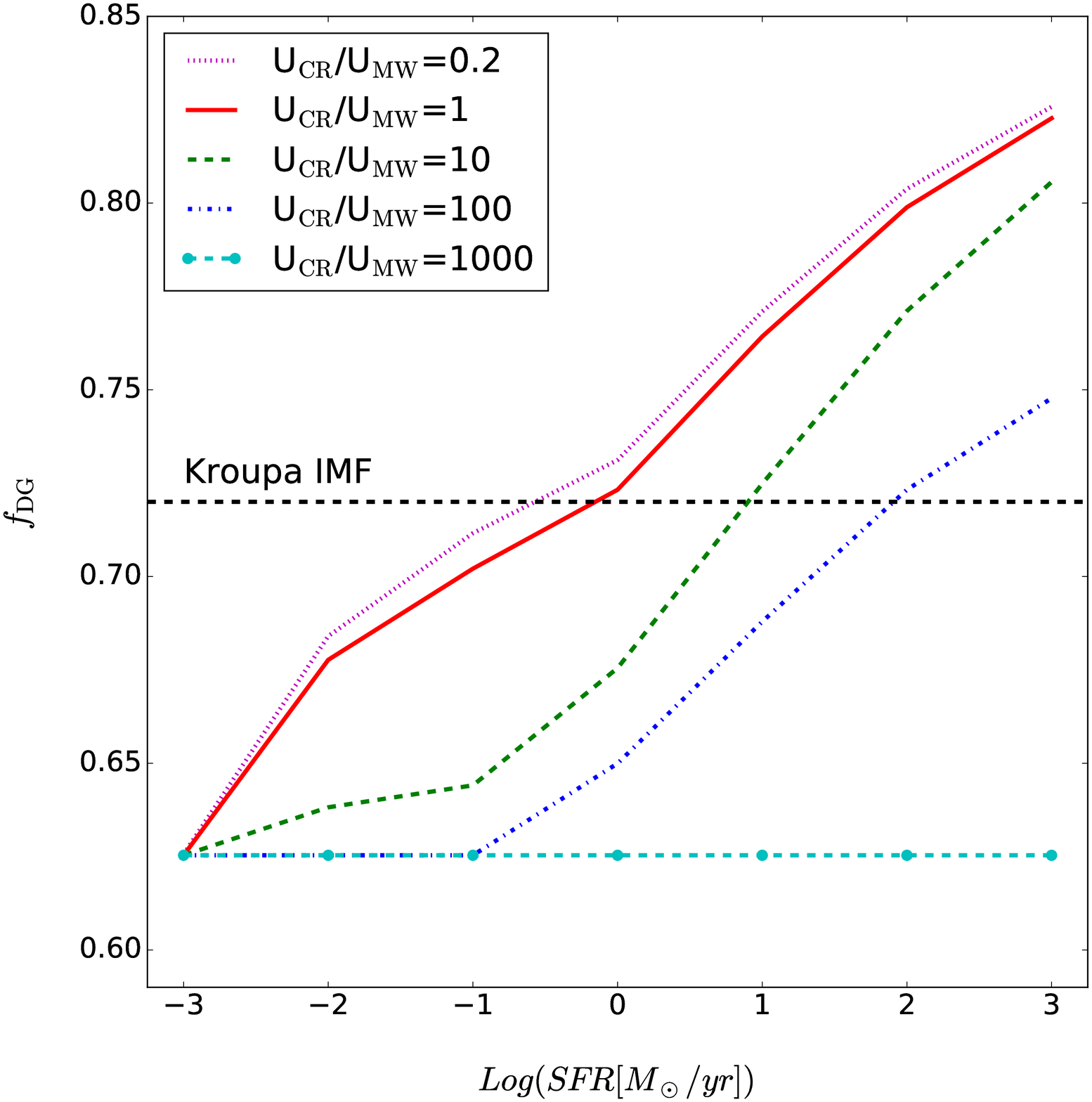} }
  \caption{Fraction of dwarf-to-giant stars, $f_{\rm dg}$, for
    individual CR-regulated IGIMFs (see the text for details).
    Different lines correspond to different CR densities, as labelled
    in the upper--left corner of the Figure. The horizontal black
    dashed line shows the $f_{\rm dg}$ value for a MW-like Kroupa IMF.
  }\label{fig:single_fdg}
\end{figure}

We can now compute the IGIMF shape as a function of both SFR and
$U_{\rm CR}/U_{\rm MW}$. It is important to keep in mind that the
original PP11 approach focuses on individual MCs, and SFR and $U_{\rm
  CR}/U_{\rm MW}$ are treated as independent variables ($U_{\rm CR}$
represents an external field). Our combined approach, instead,
considers both SFR and $U_{\rm CR}$ as ``global'' variables, given by
an average over SF regions. Therefore, in our implementation, SFR and
$U_{\rm CR}/U_{\rm MW}$ are not completely independent
variables\footnote{F18 assumed that $U_{\rm CR}$ is homogeneous within
  model galaxies, and proportional to the total disc SFR surface
  density ($\Sigma_{\rm SFR}$). As shown in Fig.~3 of F18, galaxies
  populate well defined regions in the SFR-$\Sigma_{\rm SFR}$
  space.}. In Fig.~\ref{fig:compimf} we show some representative IMF
shapes. Each panel refers to a different value of $U_{\rm CR}/U_{\rm
  MW}$, and three different SFR values (100, 1, and 0.01 $\msunyr$),
each normalized to $1 \msun$. Fig.~\ref{fig:compimf} shows that the
CR-IGIMF framework features a variety of IMF shapes, while for a SF
galaxy with MW-like energy density field it predicts an IMF shape that
is in good agreement with a Kroupa-like IMF (green solid line in the
middle-upper panel). At fixed $U_{\rm CR}/U_{\rm MW}$, the high-mass
end becomes shallower at increasing SFR, as in the original IGIMF
approach. We thus expect that a model implementing the CR-IGIMF should
be able to reproduce the observed trend in the [$\alpha$/Fe]-stellar
mass relation of ETGs \citep[see e.g.][]{Thomas10}. At the low-mass
end, the situation is more complex, and depends on $U_{\rm CR}$. For
CR densities much higher than in the MW, the low-mass end slope is
constant, whereas for $U_{\rm CR}/U_{\rm MW} \lesssim 10$ there is a
clear trend for a steepening of the IMF at increasing SFR. By
construction (see above), the minimum and maximum slopes for the
low-mass end of the CR-IGIMF are $\alpha_1$ and $\alpha_2$. In our
approach, $\alpha_1$ and $\alpha_2$ assume a fixed value ($1.3$ and a
Salpeter-like slope of $2.35$). This is a conservative choice, which
complies with both local observations of individual clouds and with
theoretical calculations of the fragmentation of giant MCs
\citep{HennebelleChabrier08}. However, we note that there is no
physical reason for these slopes to be the same for MCs evolving in
physical conditions very different from those in the MW. Indeed,
variations in the IMF shape of individual MCs have been considered in
the original IGIMF framework \citep[see e.g.][]{Yan17}. In particular,
Eq.~\ref{eq:alpha3} follows an empirical calibration connecting the
IMF high-mass end slope and the observed properties of local MCs
\citep{Marks12}. Nonetheless, as shown explicitly in
\citet{Papadopoulos11}, the impact of $U_{\rm CR}$ on $M^{\star}_{\rm
  J}$ is such that it can also affect the IMF shape of individual
clouds for $m_\star \lesssim 1 \msun$. Our approach accounts for this
effect, thus assuming that CRs can affect the IMF shape on a wider
stellar mass range than in the original IGIMF formulation.
\begin{figure}
  \centerline{ \includegraphics[width=9cm]{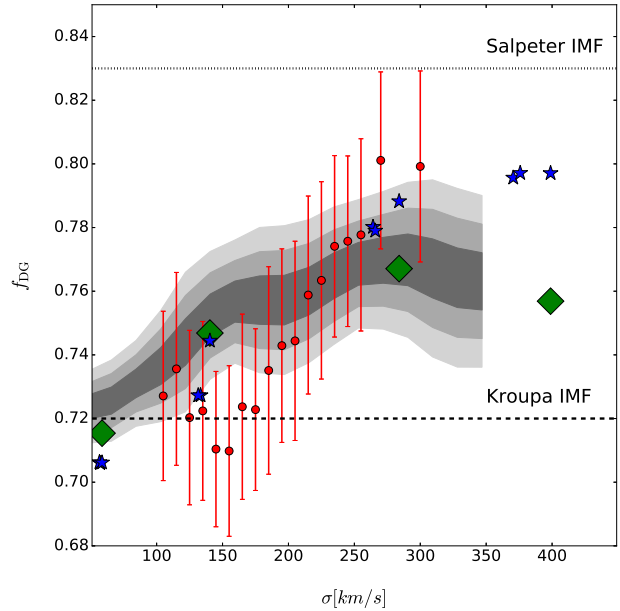} }
  \caption{The expected fraction of dwarf-to-giant stars, $f_{\rm
      dg}$, is plotted against galaxy velocity dispersion, for mock
    SFHs extracted from semi-analytic models. Red bullets with
    errorbars show the observational constraints from LB13, after
    correcting $f_{\rm dg}$ to an aperture of one effective radius by
    assuming IMF radial gradients as in~\citet{LaBarbera17}. Blue
    stars are predicted values of $f_{\rm dg}$ for mean SFHs extracted
    from {\gaea}, in reference stellar mass bins (i.e. $M_{\star} \sim
    10^{12}$, $10^{11.5}$, $10^{10.5}$, and $10^{9.25} \msun$,
    respectively), and assuming a fixed $U_{\rm CR}/U_{\rm
      MW}=1$. Grey shaded areas correspond to 1-, 2- and 3-$\sigma$
    levels around the mean $f_{\rm dg}$ versus $\sigma$ relation for
    the mocks galaxy sample extracted from the PP11 run, while green
    diamonds correspond to the predictions for mean SFHs in our
    reference mass bins (see the text for details).}\label{fig:fdg}
\end{figure}
\begin{figure}
  \centerline{ \includegraphics[width=9cm]{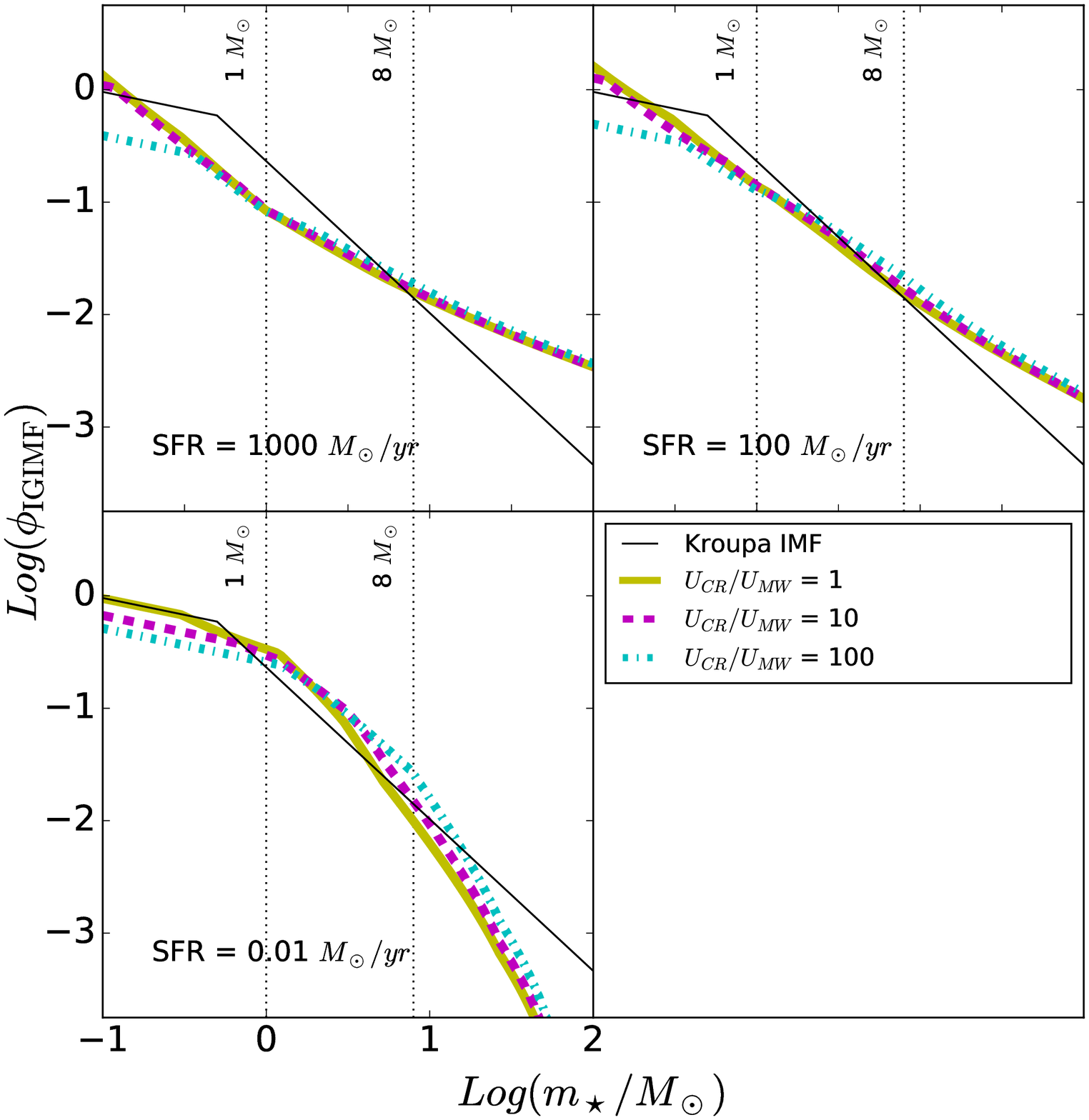} }
  \caption{Same as in Fig.~\ref{fig:compimf}, but different panels
    refer to different SFR levels, as labelled in the lower--left
    corner of each panel. Yellow solid, magenta dashed and cyan dotted
    lines correspond to $U_{\rm CR}/U_{\rm MW} = $1, 10, 100,
    respectively. In all panels, the thin solid line shows the
    canonical \citet{Kroupa01} IMF.}\label{fig:compimf_sfr}
\end{figure}

For a MW-like $U_{\rm CR}$ environment, high-SFR events correspond to
a steeper low-mass end slope and to a shallower high-mass end slope
with respect to a Kroupa IMF, {\it at the same time}. Therefore, our
implementation of the CR-IGIMF should explain the enhanced fraction of
low-mass-to-giant stars inferred from IMF-sensitive features in the
spectra of ETGs~\citep{ConroyvanDokkum12, LaBarbera17, Sarzi17}, while
qualitatively preserving most of the results discussed in F17 and
F18. Analysing these aspects in detail requires a self-consistent
theoretical model including explicitly the effects of the CR-IGIMF. In
the present work, we present a preliminary analysis focusing on the
mass fraction of low-mass stars ($f_{\rm dg}$), defined as the fraction
of mass in stars with $m_\star < 0.6 \msun$, with respect to the total
mass in stars $m_\star < 1.0 \msun$.

The value $f_{\rm dg}$ represents the dwarf-to-giant ratio, that is
the main quantity constrained by IMF-sensitive spectral features in
the spectra of $z \sim 0$ ETGs. Using a large SDSS sample of ETGs,
\citet{Ferreras13} and~LB13 constructed 18 stacked galaxy spectra,
covering a velocity dispersion range from $\sim 100$ to $\rm \sim 300
\, km/s$. The authors found a systematic steepening of the low-mass
end as a function of galaxy velocity dispersion, and parameterized the
IMF by either a single power-law (``unimodal'') or a low-mass tapered
(``bimodal'') distribution. We consider here results of the
``2SSP+XFe'' fitting method of LB13, where spectral indices are fitted
with stellar population models including two single stellar
populations (SSPs), and accounting for the effect of non-solar
abundance ratios. As shown in LB13, both unimodal and bimodal models
fit equally well the data, but provide very different mass-to-light
ratio estimates. The two IMF parameterizations provide, however, very
similar dwarf-to-giant fractions. Computing $f_{\rm dg}$ from the LB13
``2SSP+XFe'' results, we find values ranging between 0.6 and
0.95. These values refer to the inner core regions of ETGs, as
observed by the SDSS fiber aperture spectra. For galaxies in the
highest mass bin considered in LB13, this aperture corresponds to
about $1/4$ of a galaxy effective radius. We compute $f_{\rm dg}$ for
individual CR-IGIMFs, using a grid of 42 independent realizations,
with seven SFR levels (with $\log(SFR/\msunyr) = $, -3, -2, -1, 0, 1,
2, 3) and six $U_{\rm CR}/U_{\rm MW}$ values (0.2, 1, 10, 100, 1000,
10000, as in \citealt{Papadopoulos11}). On this CR-IGIMF grid, $f_{\rm
  dg}$ ranges between 0.6 and 0.8, as shown in
Fig.~\ref{fig:single_fdg}. The reason for this range of values is
that, by construction (see Eq.~\ref{eq:kroimf}), no CR-IGIMF in our
approach can have a slope steeper than $\alpha_2 = 2.35$ (i.e.  a
Salpeter IMF), or shallower than $\alpha_1 = 1.3$. For single
power-law IMFs, these slopes correspond to $f_{\rm dg}=0.84$
($\alpha_2$) and 0.6 ($\alpha_1$), respectively. The Figure shows a
clear trend for an increase of $f_{\rm dg}$ with SFR at fixed $U_{\rm
  CR}<100$. Viceversa, $f_{\rm dg}$ decreases with increasing
$U_{\rm}$, at fixed SFR. In particular, for CR density values larger
than 100 times that of the MW, the low-mass end slope of the IMF is
constant, with ratios below the expectation for a Kroupa IMF. In order
to get $f_{\rm dg}$ values as high as 0.95 (found for the core regions
of the most massive ETGs in the LB13 sample), one should postulate
that also the intrinsic IMF for individual MCs should assume
``super''-Salpeter slopes under specific physical conditions (e.g.
for high- metallicity/density environments;
see~\citet{MartinNavarro15}).

\section{Results and Discussion}\label{sec:final}
The crucial aspect we want to test in the present work is if the
CR-IGIMF scenario, once embedded in a realistic galaxy formation
context, can match the observed trends. To this purpose, we use mean
star formation histories (SFHs) predicted by the {\gaea} semi-analytic
model \citep{Hirschmann16} for $z=0$ galaxies in four different
stellar mass bins, namely $M_{\star} \sim 10^{12}$, $10^{11.5}$,
$10^{10.5}$, and $10^{9.25} \msun$, respectively. These SFHs have been
extracted from the reference {\gaea} realisation, and from runs
implementing variable IMF approaches (F17 and F18). Galaxies more
massive than $\sim10^{10.5}$ are typically characterized by an early
peak of SFR, followed by a smooth decay. The epoch, height and width
of the peak depend on the final galaxy stellar mass, with more massive
galaxies having an earlier, narrower, and higher peak \citep[see
  also][]{DeLucia06}. Lower mass galaxies have almost constant SFHs.
For the mean SFH in each mass bin, we compute the mass-weighted global
IMF using our CR-IGIMF grid, and then the corresponding value of
$f_{\rm dg}$. It is worth noting that in our estimate for $f_{\rm dg}$
we only consider the $<1 M_\odot$ mass range, where the mass weighted
IMF coincides with the present day mass function. Moreover,
observational constraints are closer to luminosity-weighted
IMFs. However, for very old stellar populations (i.e. massive ETGs),
we do not expect this difference to change our conclusions
significantly. In Fig.~\ref{fig:fdg}, we compare these $f_{\rm dg}$
for $U_{\rm CR}/U_{\rm MW}=1$ (blue stars - for each mass bin we plot
separately the values corresponding to the mean SFHs extracted from
the three {\gaea} realizations considered), with the observed
values. We correct the observed values of $f_{\rm dg}$ from LB13 to an
aperture of one effective radius, whose properties we consider to be
comparable to SAM predictions. To this aim, we assume an IMF radial
profile with the same shape as that recently derived for one massive
ETG by~\citet{LaBarbera17}. For each mass bin, the profile is
rescaled, in the central region, in order to match the $f_{\rm dg}$
value within the SDSS fiber aperture, as detailed in LB13. The
correction, which is larger at highest mass, and negligible at lowest
mass~\footnote{ It is worth noting that this approach is consistent
  with recent findings that lowest mass galaxies have shallower
  gradients compared to the most massive ones \citep{MartinNavarro15,
    Parikh18}.}, brings all $f_{\rm dg}$ values from LB13 in the range
from $\sim 0.72$ to $\sim 0.8$ as illustrated in Fig.~\ref{fig:fdg}
(see red dots with error bars, corresponding to the 18 stacked spectra
of LB13). Remarkably, the aperture correction brings the observed
$f_{\rm dg}$ values within the range predicted by the CR-IGIMF.

For model galaxies, we estimate $\sigma$ using the stellar
mass-$\sigma$ relation derived by \citet{Zahid16}, and compute stellar
masses from the SFHs under the hypothesis of a Kroupa IMF,
consistently with the adopted mass-$\sigma$ relation. We checked that
our conclusions hold if we estimate the actual stellar mass associated
with the adopted SFHs, i.e.  using the actual IMF shape and
appropriate mass fraction locked into stellar remnants for each
varying IMF. Model predictions with $U_{\rm CR}/U_{\rm MW}=1$
reproduce well the trend of increasing $f_{\rm dg}$ with $\sigma$. We
verified that this applies also to models with fixed $U_{\rm
  CR}/U_{\rm MW} \lesssim 10$. We finally relax the hypothesis of a
uniform $U_{\rm CR}/U_{\rm MW}$ for our toy SFHs, and we consider
predictions from the {\gaea} run implementing the PP11 approach (as
defined in F18). In this run, we are able to track at the same time
the evolution of SFR and $\Sigma_{\rm SFR}$ for each model galaxy. As
in F18, we assume $U_{\rm CR}/U_{\rm MW} = \Sigma_{\rm
  SFR}/\Sigma_{\rm MW}$, i.e.  that the SFR density is a good proxy
for $U_{\rm CR}$ over the star-forming disc. We then compute the mean
SFRs and SFR densities as a function of cosmic time in our reference
mass bins. We also extract individual SFHs and SFR density histories
for $\sim$380000 individual $z=0$ model galaxies from the same
realisation. This information allows us to follow the time evolution
of our model galaxies in the CR-IGIMF library. Fig.~\ref{fig:fdg}
shows the resulting $f_{\rm dg}$ as a function of $\sigma$. Green
diamonds mark average values corresponding to the four reference mass
bins. The grey areas correspond to the 1-, 2-, 3-$\sigma$ levels
around the mean relation for the mock galaxies sample extracted from
the the PP11 run, that nicely match the (aperture-corrected) trend of
increasing $f_{\rm dg}$ with $\sigma$ obtained by LB13. Interestingly,
the trend is predicted to flatten at the highest stellar mass bin
probed by the F18 models, although the number of mock galaxies in this
range is small. This is mainly due to the fact that more massive
galaxies in F18 form preferentially in strong compact starburst, with
high SFR and $\Sigma_{\rm SFR}$, that correspond to lower $f_{\rm
  dg}$. To better highlight this effect, we show in
Fig.~\ref{fig:compimf_sfr} the different CR-IGIMF shapes at fixed SFR,
as a function of $U_{\rm CR}/U_{\rm MW}$. In our toy SFHs a massive
galaxy is characterized by an early large peak of SF: in these
conditions (upper panels in Fig.~\ref{fig:compimf_sfr}) our CR-IGIMF
scenario consistently predicts a shallower high-mass end slope. The
low-mass end slope, that is mainly responsible for setting $f_{\rm
  dg}$ strongly depends on the SFR density (that we use as a proxy for
$U_{\rm CR}$). Therefore, an accurate estimate for the CR energy
density is a crucial element to correctly predict the shape of the
CR-IGIMF. The late-time evolution of a massive galaxy is associated
with low-SFR and low-SFR density, that correspond to shallower
low-mass end slopes (bottom panel of Fig.~\ref{fig:compimf_sfr}).

\begin{figure}
  \centerline{ \includegraphics[width=9cm]{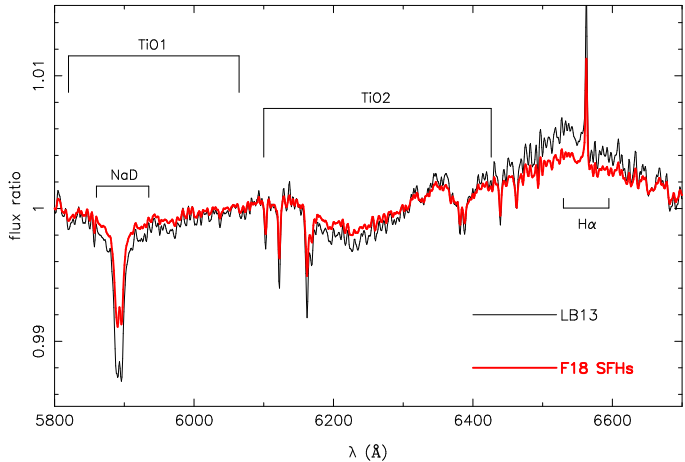} }
  \caption{Thick red line shows the ratio between toy model spectra
    corresponding to the $M_{\star}=10^{9.25}$ and $10^{11.5} \,
    M_{\odot}$ F18 galaxies, in the CR-IGIMF scenario.  This is
    compared to the observed IMF variation signal in similar mass
    bins, i.e. the ratio of {\sc MILES} SSP models that better
    describe observations from LB13, after applying an aperture
    correction, as detailed in the text. Horizontal brackets, with
    labels, mark several IMF-sensitive features in the plotted
    spectral range.}\label{fig:IMFresponse}
\end{figure}

In order to further compare predictions from our CR-IGIMF model to
observations, we compute synthetic model spectra corresponding to the
F18 mean SFHs in the reference mass bins of $\rm M_{\star} =
10^{9.25}$ and $10^{11.5}$ ($\sigma \sim 60$ and $\rm \sim 280 \,
km/s$). We consider the {\sc MILES}\footnote{miles.iac.es} stellar
library and the Padova 2000 isochrones \citep{Girardi00}, and we build
up the spectral energy distributions (SEDs) by summing up the SSPs
corresponding to the mean (mass-weighted) IMF in the CR-IGIMF
framework\footnote{In order to obtain the desired values of
  metallicity, gravity, and effective temperature, we interpolate {\sc
    MILES} spectra using the same algorithm as in
  \citet{Vazdekis10}. We then weight the interpolated spectra using
  the mass--(V-band)luminosity relation, rather than using empirical
  relations (as in \citealt{Vazdekis10}). Therefore, the resulting
  SSPs should be regarded as simple toy models, and not full MILES SSP
  SEDs.}. Since we want to highlight the effect of a varying IMF,
rather than other stellar population properties, for both mass bins,
we consider the same (old) age of $\sim$10 Gyr, and solar
metallicity. Fig.~\ref{fig:IMFresponse} shows, as thick red curve, the
ratio between these toy model spectra for the highest relative to
lowest mass bin. This ratio represents the IMF signal corresponding to
the CR-IGIMF implementation applied to F18 mean SFHs. The Figure plots
the wavelength range from $5800$ to $\sim 6700$~\AA, where several
IMF-sensitive feature (like TiO1 and TiO2) are located (as marked in
the plot). The thin black line in Fig.~\ref{fig:IMFresponse}
represents the ``observed'' IMF signal, obtained by dividing a bimodal
SSP {\sc MILES} model~\citep{Vazdekis10} with slope $\Gamma_b=1.8$
(that should well describe massive galaxies), to that for a Kroupa IMF
(that well describes the lowest mass stacks of LB13). In deriving the
thin black line, we account for aperture correction in real data and
we consider SSPs with the same age and metallicity as those used for
the mean F18 SFHs. Fig.~\ref{fig:IMFresponse} shows that the CR-IGIMF
and ``observed'' IMF variations from low- to high-mass galaxies, agree
reasonably well, consistent with what seen in Fig.~\ref{fig:fdg}, for
the dwarf-to-giant ratio, $f_{\rm dg}$.

The results shown in Figs.~\ref{fig:fdg} and~\ref{fig:IMFresponse} are
based on SFHs extracted from {\gaea} runs, hence implementing a
different varying-IMF approach than the CR-IGIMF framework described
here. In F18, we have shown that a varying-IMF scenario can
significantly affect galaxy evolution, as predicted by {\gaea}, since
it changes the IMF that we associate to each individual SF event
predicted by the model. A model implementing a variable IMF such as
the CR-IGIMF thus implies a time variation of the IMF: this effect has
deep implications for the predicted SFHs of the different galaxy
populations \citep[see e.g.][]{FerreMateu13}. Therefore, the
development of a self-consistent version of {\gaea}, implementing the
CR-IGIMF, is desirable, and will be presented in a future work. Such a
model would allow us to carry out a more detailed comparison with
observational data, including a direct comparison of synthetic spectra
for our mock galaxies with observed ones. Another aspect that deserves
further investigation is that of IMF radial gradients in ETGs. We
expect the central regions (i.e. those showing a stronger signal for a
``bottom-heavy'' IMF) to have larger SFRs and SFR densities, with
respect to the galaxy outskirts. Naively, in the current CR-IGIMF
implementation, we would tend to associate a lower $f_{\rm dg}$ to
these regions, in contrast to some recent observational
results~\citep{MartinNavarro15, LaBarbera16, vanDokkum17,
  Sarzi17}. Again, a self-consistent version of {\gaea} implementing
the CR-IGIMF is clearly needed to study this aspect, taking into
account the different star-forming regions associated with the
hierarchical assembly of massive ETGs and their individual
IMFs. Moreover, as mentioned above, one should consider that specific
physical properties (e.g. metallicity - see~\citealt{MartinNavarro15})
may affect the IMF of individual MCs in the core regions of massive
ETGs, hence requiring further modifications to the CR-IGIMF framework.

The results presented in this work are meant to show the capabilities
of the CR-IGIMF approach in a galaxy evolution framework. Our models
predict $f_{\rm dg}$ values, over a spatial scale comparable to that
of galaxy sizes, that are consistent with the observed ones. This
demonstrates that the CR-IGIMF scenario is able to capture key
features of the IMF low- and high-mass end, that have been elusive so
far. Forthcoming work will focus on the development of theoretical
models self-consistently including both the effect of CR-IGIMF on
model galaxy evolution and the synthesis of realistic mock
spectra. The present work shows that these tools are of fundamental
importance for a interpreting the wealth of observational data in
favour of a variable IMF.

\section*{Acknowledgements}

FF thanks P. Kroupa for stimulating discussions on the definition of
the IGIMF framework. GDL thanks the Alexander von Humboldt Foundation
and the Astronomisches Rechen-Institut of Heidelberg for the pleasant
and productive stay during which part of this work was carried
out. FLB and AV acknowledge support from grant AYA2016-77237-C3-1-P
from the Spanish Ministry of Economy and Competitiveness
(MINECO). Some figures have been draw thanks to the {\sc pygtc}
package \citep{Bocquet2016}.

\bibliographystyle{mn2e}
\bibliography{fontanot}

\end{document}